\UseRawInputEncoding
\documentclass[twocolumn,showkeys,aps,prb,showpacs]{revtex4-1}
\usepackage{graphicx}
\usepackage[CJKbookmarks,dvipdfm,colorlinks,linkcolor=blue,citecolor=blue]{hyperref}

\begin{document}

\title{Intrinsic room-temperature piezoelectric  quantum anomalous hall insulator in Janus monolayer  $\mathrm{Fe_2IX}$ (X=Cl and Br)}

\author{San-Dong Guo$^{1}$, Wen-Qi Mu$^{1}$,  Xiang-Bo Xiao$^{2,3}$  and  Bang-Gui Liu$^{2,3}$}
\affiliation{$^1$School of Electronic Engineering, Xi'an University of Posts and Telecommunications, Xi'an 710121, China}
\affiliation{$^2$ Beijing National Laboratory for Condensed Matter Physics, Institute of Physics, Chinese Academy of Sciences, Beijing 100190, People's Republic of China}
\affiliation{$^3$School of Physical Sciences, University of Chinese Academy of Sciences, Beijing 100190, People's Republic of China}
\begin{abstract}
A  two-dimensional (2D) material  with piezoelectricity, topological and ferromagnetic (FM) orders, namely 2D piezoelectric quantum anomalous hall insulator (PQAHI), may open new opportunities to realize novel physics and applications.
Here, by first-principles calculations,
a family of  2D Janus monolayer  $\mathrm{Fe_2IX}$ (X=Cl and Br) with  dynamic, mechanical and thermal  stabilities is  predict to be  room-temperature PQAHI. At the absence of spin-orbit coupling (SOC),  monolayer  $\mathrm{Fe_2IX}$ (X=Cl and Br) is  a half Dirac semimetal state. When  the SOC is included, these monolayers become quantum anomalous hall  (QAH) states with
sizable gaps (more than two hundred meV) and two chiral edge modes (Chern number C=2). It is also found that monolayer $\mathrm{Fe_2IX}$ (X=Cl and Br) possesses robust QAH states against biaxial strain.  By symmetry analysis, it is found that only out-of-plane piezoelectric response can be
induced by a uniaxial strain in the basal plane.  The calculated  out-of-plane $d_{31}$ of  $\mathrm{Fe_2ICl}$ ($\mathrm{Fe_2IBr}$) is 0.467 pm/V  (0.384 pm/V), which  is  higher than  or comparable with ones of  other 2D known materials.
Meanwhile, using Monte Carlo (MC)
simulations, the Curie temperature $T_C$ is estimated to be 429/403 K  for monolayer  $\mathrm{Fe_2ICl}$/$\mathrm{Fe_2IBr}$ at FM ground state, which is above room temperature. Finally, the interplay of electronic correlations with nontrivial
band topology is studied to confirm the robustness of QAH state.
The combination of piezoelectricity, topological and FM orders makes monolayer  $\mathrm{Fe_2IX}$ (X=Cl and Br) become a potential platform for multi-functional spintronic applications with large gap and high $T_C$.
Our works provide possibility to use the piezotronic effect to control QAH effects, and  can stimulate further experimental works.

\end{abstract}

\pacs{71.20.-b, 77.65.-j, 72.15.Jf, 78.67.-n ~~~~~~~~~~~~~~~~~~~~~~~~~~~~~~~~~~~Email:sandongyuwang@163.com}

\maketitle

\section{Introduction}
Seeking 2D multifunctional piezoelectric  materials is a compelling problem of novel physics and materials science\cite{z}, like the combination of
piezoelectricity with topological
insulating phase, intrinsic ferromagnetism and quantum
spin hall effect.  Using piezoelectric effect to control the quantum or spin transport process may lead to
novel device applications or scientific breakthroughs.
Both in experiment and in theory, the  advances have been made on 2D piezoelectric  materials.
Experimentally, the  piezoelectricity has been reported in monolayer   $\mathrm{MoS_2}$, MoSSe and $\mathrm{In_2Se_3}$\cite{q5,q6,q8,q8-1}.
The piezoelectric properties of many 2D materials  without inversion symmetry have been investigated by density functional theory (DFT) calculations, including strain-tuned effects on piezoelectricity\cite{q7-0,q7-1,q7-2,q7-4,q7-7,q7-8,q7-10,q9-0,q9-1,q9}.
Recently, some progress have been made on 2D multifunctional piezoelectric  materials.
The coexistence of intrinsic piezoelectricity and ferromagnetism, namely  piezoelectric ferromagnetism (PFM), has been predicted in 2D vanadium dichalcogenides,  $\mathrm{VSi_2P_4}$, $\mathrm{CrBr_{1.5}I_{1.5}}$ and $\mathrm{InCrTe_3}$\cite{qt1,q15,q15-1,q15-2}.
The combination of piezoelectricity with   topological
insulating phase has also been  achieved in monolayer  InXO (X=Se and Te)\cite{gsd1} and Janus monolayer $\mathrm{SrAlGaSe_4}$\cite{gsd2}.
A natural idea is to search for  PQAHI with piezoelectricity, topological and FM orders.

\begin{figure*}
  \includegraphics[width=12.0cm]{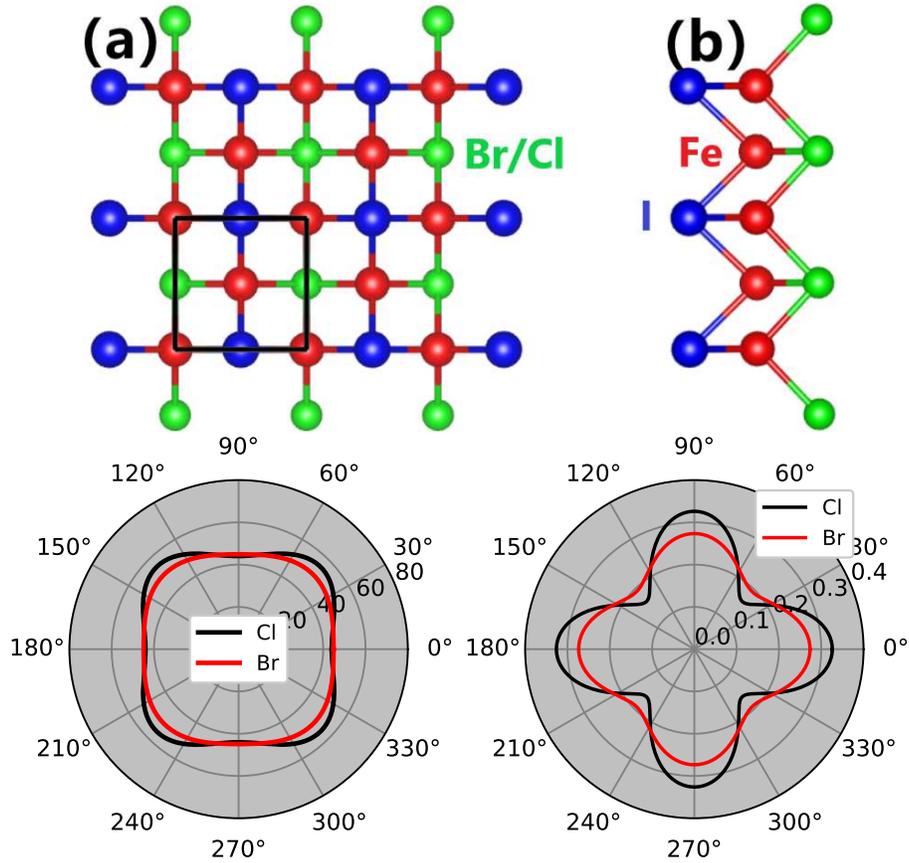}\\
  \caption{(Color online)Top:the (a) top view and (b) side view of  crystal structure of Janus monolayer $\mathrm{Fe_2IX}$ (X=Cl and Br).  The  black frame represents the  primitive cell. Bottom:the angular dependence of the Young's modulus ($C_{2D}(\theta)$) and Poisson's
ratio ($\nu_{2D}(\theta)$) of Janus monolayer  $\mathrm{Fe_2IX}$ (X=Cl and Br).}\label{t0}
\end{figure*}

\begin{figure}
  \includegraphics[width=8cm]{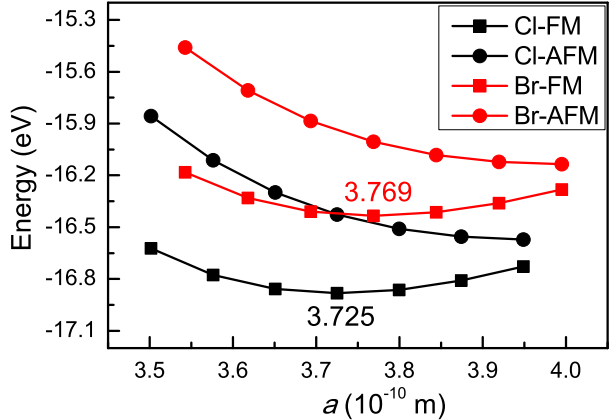}
\caption{(Color online) The FM and AFM energy of  Janus monolayer  $\mathrm{Fe_2IX}$ (X=Cl and Br) as a function of lattice constants $a$. }\label{e}
\end{figure}

Many kinds of 2D magnetic materials have been widely  investigated\cite{m7-1,m7-2,m7-3,m7-4,m7-6,m7-7,m7-8}, and   the monolayer $\mathrm{Cr_2Ge_2Te_6}$, $\mathrm{CrI_3}$, $\mathrm{VS_2}$ and $\mathrm{VSe_2}$ have  been experimentally confirmed\cite{m7-2,m7-4,m7-6}.
The  combination of ferromagnetism with   topological
insulating phase will  produce QAH insulator, which can be characterized by a nonzero
Chern number, being in accordance with the number of edge states.
The QAH insulator is first observed
experimentally  in thin films of Cr doped
$\mathrm{(Bi, Sb)_2Te}$ below 30 mK\cite{qa1}. Although various
theoretical predictions and the new synthetic
methods have been proposed,  a higher-temperature
QAH insulator is still a  challenge in experiment\cite{qa1,qa2}.
Recently, a
robust QAH insulator  $\mathrm{Fe_2I_2}$ monolayer is proposed with  a topologically nontrivial band gap of
301 meV, a nonzero Chern number C=2 and a
high Curie temperature (about 400 K)\cite{fe}.  The experimental feasibility and stability of
$\mathrm{Fe_2I_2}$ monolayer have been proved by the cohesive energy, phonon spectra and ab initio
molecular dynamics (AIMD) simulations. However, $\mathrm{Fe_2I_2}$ monolayer possesses no piezoelectricity due to inversion symmetry.

 It's a natural idea to achieve PQAHI, based on monolayer $\mathrm{Fe_2I_2}$.
It is noted that the monolayer $\mathrm{Fe_2I_2}$  has  sandwiched I-Fe-I structure. Inspiring from the already synthesized Janus monolayer MoSSe\cite{p1}, constructed  by  replacing one of two  S   layers with Se  atoms in  $\mathrm{MoS_2}$ monolayer,  it is possible to achieve Janus structure based on   $\mathrm{Fe_2I_2}$ monolayer, and then produce  piezoelectricity.
In this work,    Janus monolayer  $\mathrm{Fe_2IX}$ (X=Cl and Br) is constructed based on QAH insulator  $\mathrm{Fe_2I_2}$ monolayer by replacing the top I atomic layer  with X atoms. These Janus monolayer are  dynamically, mechanically  and thermally stable.
By first-principles
calculations, the nontrivial topological state of monolayer  $\mathrm{Fe_2IX}$ (X=Cl and Br)  is
firmly confirmed by a nonzero Chern number (C=2) and
chiral edge states. Their  nontrivial band gaps are larger than 200
meV, and the Curie temperatures are  estimated
to be about 400 K.  By symmetry analysis, only out-of-plane piezoelectric response can be
induced by a uniaxial strain, and the predicted  out-of-plane $d_{31}$ of  $\mathrm{Fe_2ICl}$ ($\mathrm{Fe_2IBr}$) is 0.467 pm/V  (0.384 pm/V), which  is  higher than  or comparable with ones of  other 2D known materials.
These  indicate the enormous potential of Janus monolayer  $\mathrm{Fe_2IX}$ (X=Cl and Br)
in developing 2D piezoelectric spin topological devices.

The rest of the paper is organized as follows. In the next
section, we shall give our computational details and methods.
 In  the next few sections,  we shall present crystal structure, structural stabilities,  topological properties,   piezoelectric properties and Curie temperatures of Janus monolayer  $\mathrm{Fe_2IX}$ (X=Cl and Br).  Finally, we shall give our discussion and conclusions.

\begin{figure}
  \includegraphics[width=8cm]{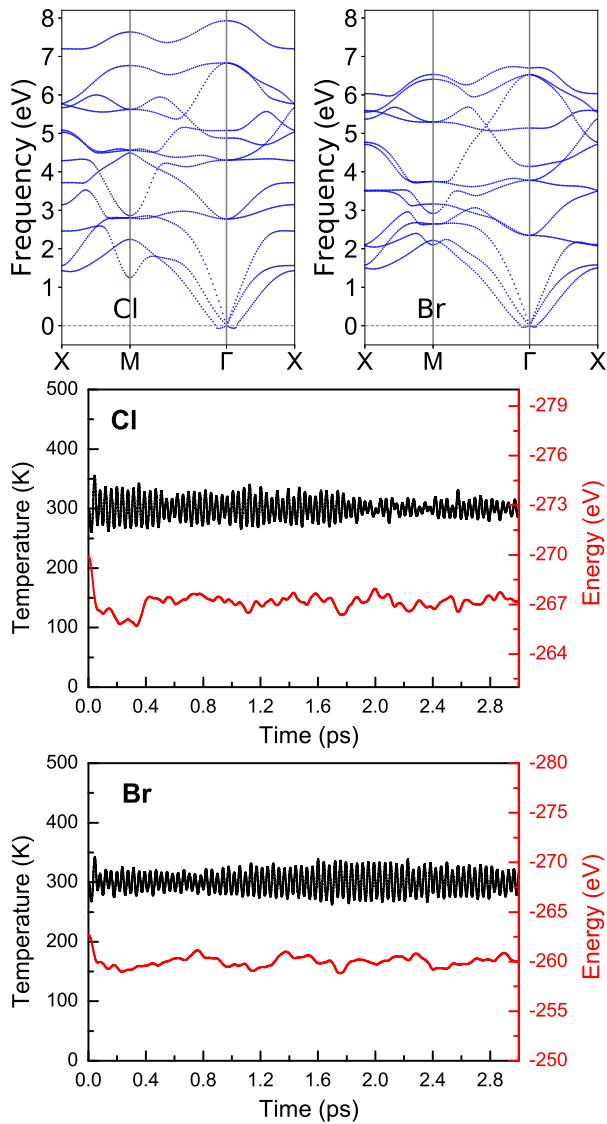}
  \caption{(Color online)Top:the phonon band dispersions  of Janus monolayer $\mathrm{Fe_2IX}$ (X=Cl and Br)  with FM  magnetic configuration. Bottom:the temperature and total energy fluctuations of  Janus monolayer  $\mathrm{Fe_2IX}$ (X=Cl and Br)  with FM  magnetic configuration at 300 K.}\label{phon}
\end{figure}

\section{Computational detail}
First-principles calculations are carried out  by DFT\cite{1} using the VASP code\cite{pv1,pv2,pv3}. The plane-wave basis with an energy cutoff of 500 eV is used for Janus monolayer  $\mathrm{Fe_2IX}$ (X=Cl and Br).
The exchange-correlation functional is approximated by the popular generalized gradient approximation (GGA) of Perdew, Burke and  Ernzerhof\cite{pbe}. The SOC is included in the calculations self-consistently. The total energy  convergence criterion is set  for $10^{-8}$ eV.  All the lattice constants and atomic coordinates are optimized with  the force on each atom being  less than 0.0001 $\mathrm{eV.{\AA}^{-1}}$.  A vacuum spacing of more than 15 $\mathrm{{\AA}}$ is used  to avoid artificial interactions caused by the periodic boundary condition. The
DFT+$U$ method\cite{u} is employed for the treatment of the strongly
correlated $3d$ electrons of Fe atom with $U_{eff}$ = 2.5 eV\cite{fe,fe1} for studied monolayers.

The elastic stiffness tensor  $C_{ij}$   are carried out by using strain-stress relationship (SSR) with GGA,  and  the piezoelectric stress tensor $e_{ij}$  are calculated  by   density functional perturbation theory (DFPT) method\cite{pv6} using GGA+SOC.
A Monkhorst-Pack k-mesh of 18$\times$18$\times$1 is used to sample the Brillouin Zone (BZ) for the self-consistent calculations   and elastic coefficients $C_{ij}$ . To attain the accurate $e_{ij}$,   a dense  mesh of 26$\times$26$\times$1 k-points is adopted.
The 2D elastic coefficients $C^{2D}_{ij}$
 and   piezoelectric stress coefficients $e^{2D}_{ij}$
have been renormalized by   $C^{2D}_{ij}$=$Lz$$C^{3D}_{ij}$ and $e^{2D}_{ij}$=$Lz$$e^{3D}_{ij}$, where the $Lz$ is  the length of unit cell along z direction.

 The interatomic force constants (IFCs) with the 5$\times$5$\times$1 supercell are obtained
 with FM ground state by finite displacement method.  From calculated  harmonic IFCs, the
phonon dispersions are evaluated
using Phonopy code\cite{pv5}. Surface state and Berry curvature calculations are carried out by WannierTools code, based on the tight-binding Hamiltonians constructed from maximally localized Wannier functions by Wannier90 code\cite{w1,w2}. The Curie temperature is estimated by  MC simulation using Mcsolver code\cite{mc}.

\section{Crystal Structure}
Similar to monolayer $\mathrm{Fe_2I_2}$\cite{fe}, Janus monolayer  $\mathrm{Fe_2IX}$ (X=Cl and Br)  contains three atomic sublayers with Fe layer sandwiched between I and X layers, whose unit cell contains four atoms with two co-planar Fe atoms. The schematic crystal structures of Janus monolayer  $\mathrm{Fe_2IX}$ (X=Cl and Br) are shown in \autoref{t0}. Similar to Janus monolayer MoSSe from $\mathrm{MoS_2}$\cite{p1}, the  Janus monolayer   $\mathrm{Fe_2IX}$ (X=Cl and Br) can be constructed  by  replacing one of two I  layers with X  atoms in monolayer  $\mathrm{Fe_2I_2}$.
 Janus monolayer  $\mathrm{Fe_2IX}$ (X=Cl and Br)  crystallizes in the orthorhombic $P4mm$ space group (No.99), which loses centrosymmetry compared to centrosymmetric monolayer  $\mathrm{Fe_2I_2}$ with $P4/nmm$ space group (No. 129). The missing centrosymmetry will lead to piezoelectricity for Janus monolayer  $\mathrm{Fe_2IX}$ (X=Cl and Br).

 To determine the ground state of  $\mathrm{Fe_2IX}$ (X=Cl and Br),  two different initial magnetic configurations, including antiferromagnetic (AFM) and FM states, are used, and  the energy of   FM  and AFM states  as a function of lattice constants $a$ are shown in \autoref{e}.
It is clearly seen  that the FM order is the most stable magnetic state for Janus monolayer  $\mathrm{Fe_2IX}$ (X=Cl and Br), which
 means that
ferromagnetism in monolayer  $\mathrm{Fe_2I_2}$ is retained by elements substitution engineering.
 The optimized lattice constants with FM state is 3.725 $\mathrm{{\AA}}$ for  $\mathrm{Fe_2ICl}$, and 3.769   $\mathrm{{\AA}}$ for  $\mathrm{Fe_2IBr}$.
 Due to different atomic sizes and electronegativities of Cl/Br and I atoms,  the inequivalent Fe-Cl/Br and Fe-I bond lengths (Cl/Br-Fe-Cl/Br and I-Fe-I bond  angles) from \autoref{tab} can induce a built-in electric field. It is clearly seen that $\mathrm{Fe_2ICl}$ has greater deviation from horizontal mirror symmetry than $\mathrm{Fe_2IBr}$.
\begin{figure}
  \includegraphics[width=8cm]{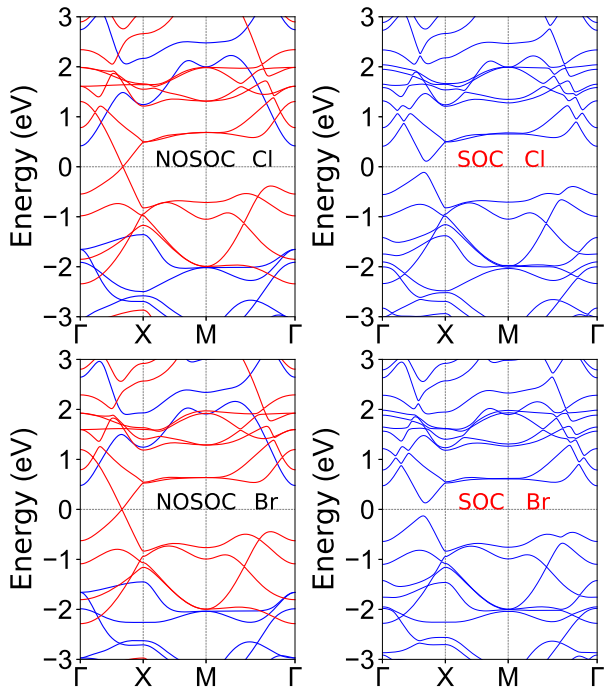}
  \caption{(Color online) The  energy band structures of  Janus monolayer  $\mathrm{Fe_2IX}$ (X=Cl and Br) without and with SOC at the FM  ground state. The blue (red) lines represent the band structure in the spin-up (spin-down) direction without SOC.  }\label{band}
\end{figure}

\begin{figure*}
  \includegraphics[width=12.0cm]{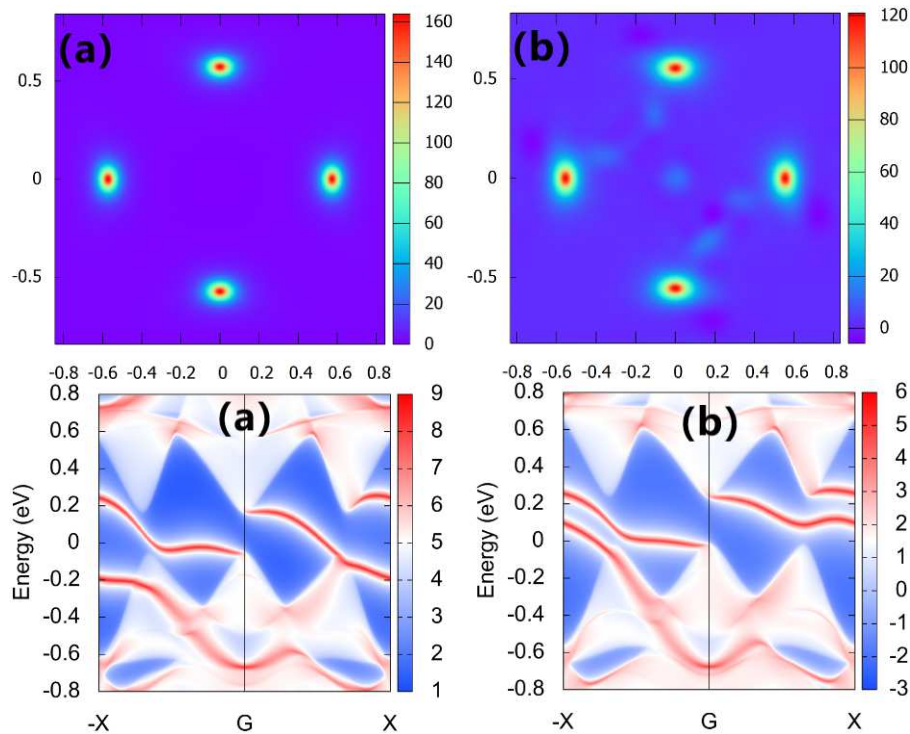}
  \caption{(Color online)Top:the distribution of Berry curvature of $\mathrm{Fe_2ICl}$ (a) and $\mathrm{Fe_2IBr}$ (b) contributed
by occupied valence bands in the momentum space. Bottom: the topological
edge states of $\mathrm{Fe_2ICl}$ (a) and $\mathrm{Fe_2IBr}$ (b) calculated  along the (100) direction.}\label{berry}
\end{figure*}

\begin{table}
\centering \caption{ The structural  parameters including lattice constants $a_0$ ($\mathrm{{\AA}}$) ;  Fe-Cl/Br ($d_1$)  and Fe-I ($d_2$) bond lengths ($\mathrm{{\AA}}$); Cl/Br-Fe-Cl/Br ($\theta_1$) and I-Fe-I ($\theta_2$) angles;  the thickness layer height ($t$) ($\mathrm{{\AA}}$). }\label{tab}
  \begin{tabular*}{0.48\textwidth}{@{\extracolsep{\fill}}ccccccc}
  \hline\hline
Name& $a_0$ &  $d_1$&$d_2$& $\theta_1$& $\theta_2$& $t$\\\hline\hline
 $\mathrm{Fe_2ICl}$&3.725&2.463&2.700&98.259&87.253&3.566\\\hline
 $\mathrm{Fe_2IBr}$&3.769&2.582&2.704&93.747&88.355&3.704\\\hline\hline
\end{tabular*}
\end{table}

\section{Structural Stability}
\autoref{phon} shows the the phonon spectra of Janus monolayer  $\mathrm{Fe_2IX}$ (X=Cl and Br), which can be used to analyze their  dynamical stabilities. The nine optical and three acoustical phonon
branches with a total of twelve branches due to four
atoms per unitcell can be observed.
 Their phonon spectra show no imaginary frequency,
which confirms  the dynamical stability of monolayer  $\mathrm{Fe_2IX}$ (X=Cl and Br), and they  can exist as free-standing
2D crystals. The thermal stability of the monolayer $\mathrm{Fe_2IX}$ (X=Cl and Br) can be checked by  AIMD simulations  using NVT ensemble with a supercell
of size 4$\times$4$\times$1 for more than
3000 fs with a time step of 1 fs.
The temperature and total energy fluctuations of monolayer $\mathrm{Fe_2IX}$ (X=Cl and Br) as a function of the simulation time are plotted in \autoref{phon} at room temperature.  No obvious structural disruption is observed with small the temperature and total energy
 fluctuates, which  confirms the thermodynamical stability of monolayer  $\mathrm{Fe_2IX}$ (X=Cl and Br) at room temperature.

The mechanical stability of monolayer  $\mathrm{Fe_2IX}$ (X=Cl and Br) can be proved by elastic
constants. For 2D materials, using Voigt notation, the elastic tensor with $4mm$ point-group symmetry  can be expressed as:
\begin{equation}\label{pe1-4}
   C=\left(
    \begin{array}{ccc}
      C_{11} & C_{12} & 0 \\
     C_{12} & C_{11} &0 \\
      0 & 0 & C_{66} \\
    \end{array}
  \right)
\end{equation}
The calculated $C_{11}$, $C_{12}$ and  $C_{66}$ are 49.28 $\mathrm{Nm^{-1}}$, 16.07 $\mathrm{Nm^{-1}}$ and 23.68 $\mathrm{Nm^{-1}}$ for  $\mathrm{Fe_2ICl}$, and 48.53  $\mathrm{Nm^{-1}}$,    13.27 $\mathrm{Nm^{-1}}$ and      20.91 $\mathrm{Nm^{-1}}$ for  $\mathrm{Fe_2IBr}$. The calculated elastic constants
satisfy the  Born  criteria of mechanical stability:
 \begin{equation}\label{pe1-4}
  C_{11}>0,~~ C_{66}>0,~~C_{11}-C_{12}>0
\end{equation}
 Therefore,  Janus monolayer  $\mathrm{Fe_2IX}$ (X=Cl and Br) is mechanically stable.

The direction-dependent mechanical properties of monolayer  $\mathrm{Fe_2IX}$ (X=Cl and Br) can be attained from the calculated $C_{ij}$.
The in-plane Young's moduli $C_{2D}(\theta)$ and
Poisson's ratios $\nu_{2D}(\theta)$ can be calculated by using the two formulas\cite{ela,ela1}:
 \begin{equation}\label{pe1-4-1}
  C_{2D}(\theta)=\frac{C_{11}C_{22}-C_{12}^2}{C_{11}m^4+C_{22}n^4+(B-2C_{12})m^2n^2}
\end{equation}
 \begin{equation}\label{pe1-4-2}
  \nu_{2D}(\theta)=\frac{(C_{11}+C_{22}-B)m^2n^2-C_{12}(m^4+n^4)}{C_{11}m^4+C_{22}n^4+(B-2C_{12})m^2n^2}
\end{equation}
where $m=sin(\theta)$, $n=cos(\theta)$ and $B=(C_{11}C_{22}-C_{12}^2)/C_{66}$. The $\theta$ is the angle of the direction with
 the x direction  as $0^{\circ}$ and the y direction as $90^{\circ}$.

The Young's moduli $C_{2D}(\theta)$ and
Poisson's ratios $\nu_{2D}(\theta)$  as a function of the angle $\theta$ are shown in \autoref{t0}.
Due to the symmetric
structure, both the Young's modulus and Poisson's ratios are
equivalent along the (100) and (010) directions, and we only consider the angle range from $0^{\circ}$ to $90^{\circ}$.  The softest
direction  is along the (100)
direction, with their Young¡¯s moduli of 44.04 $\mathrm{Nm^{-1}}$($\mathrm{Fe_2ICl}$)
and 44.90 $\mathrm{Nm^{-1}}$ ($\mathrm{Fe_2IBr}$).
The hardest direction is
 along the (110) direction, with their Young's
moduli of 54.91 $\mathrm{Nm^{-1}}$ ($\mathrm{Fe_2ICl}$) and 49.88 $\mathrm{Nm^{-1}}$ ($\mathrm{Fe_2IBr}$). The maximum value of Young's moduli is
less than that of graphene (340 $\mathrm{Nm^{-1}}$)\cite{gra}, which means extraordinary
flexibilities.
The minima of the direction-dependent Poisson's ratios  of monolayer  $\mathrm{Fe_2IX}$ (X=Cl and Br)  is  along the
(110) direction (0.160 and  0.193), while the maxima is along
the (100) direction (0.326 and  0.273).

\section{Topological properties}
The piezoelectric materials should be  semiconductors for
prohibiting current leakage. \autoref{band}  show the energy band structures of monolayer  $\mathrm{Fe_2IX}$ (X=Cl and Br) with GGA and GGA+SOC.
When the SOC is absent, monolayer  $\mathrm{Fe_2IX}$ (X=Cl and Br) gives a novel spin
polarized state with a large-gap insulator for spin up
and a gapless Dirac semimetal for spin down, namely  2D half Dirac semimetal state.
Due to similar electronic properties between  $\mathrm{Fe_2ICl}$ and  $\mathrm{Fe_2IBr}$, the element characters of the spin-up  and spin-down
bands of only $\mathrm{Fe_2IBr}$ monolayer using GGA are plotted in Fig.1 of electronic supplementary information (ESI), and the Fe $d$-orbital characters of the
minority-spin bands  are shown in Fig.2 of ESI.
Due to the short distance between Fe atoms
(about 2.65 $\mathrm{{\AA}}$), the Fe-Fe hybridization is strong  for $3d$
orbitals. For the Fe-$3d$, the spin-up channel
is fully occupied, while the partially occupied spin-down
channel crosses  the Fermi level (see Fig.1 of ESI). And then, when the magnetic moment of
neighboring Fe atoms are parallelly aligned, the direct electron
hopping is  allowed.
The strong FM kinetic exchange is in favour of FM order\cite{pr}.
It is found that the
two spin-down bands near the Fermi level are mainly
contributed by $d_{xy}$, $d_{z^2}$ and $d_{x^2-y^2}$ orbitals of Fe atom (see Fig.2 of ESI).
The special electronic structures mean that  these systems may
be  QAH insulators, when the SOC is included. In contrast to typical
Dirac cones in graphene, the four Dirac cones in monolayer  $\mathrm{Fe_2IX}$ (X=Cl and Br) are observed. When including the SOC, the Dirac gap can be produced, and the corresponding gap is 0.223 eV for $\mathrm{Fe_2ICl}$  and 0.260 eV for $\mathrm{Fe_2IBr}$. Similar results can be found in monolayer $\mathrm{Fe_2I_2}$ and lithium-decorated
iron-based superconductor materials\cite{fe,pr1}.

To verify the topologically nontrivial properties,
the Chern number of monolayer  $\mathrm{Fe_2IX}$ (X=Cl and Br) can be calculated by integrating the Berry curvature ($\Omega_z(k)$) of the
occupied bands:
 \begin{equation}\label{pe1-4}
  C=\frac{1}{2\pi}\int_{BZ}d^2k \Omega_z(k)
\end{equation}
\begin{equation}\label{pe1-4}
  \Omega_z(k)=\nabla_k\times i\langle\mu_{n,k}|\nabla_k\mu_{n,k}\rangle
\end{equation}
where $\mu_{n,k}$ is the lattice periodic part of the Bloch wave functions. The distributions of
Berry curvature in the momentum space for  $\mathrm{Fe_2IX}$ (X=Cl and Br) are plotted in \autoref{berry}.
For each gapped Dirac cone,  a quantized
Berry phase of $\pi$ can be attained from the valence states. The total Berry phase of 4$\pi$ is obtained due to four Dirac cones, which means a
high Chern number C=2. To corroborate this finding, topological
edge states  along the (100) direction are calculated, which are shown in \autoref{berry}.
It is clearly seen that two chiral gapless edge modes appear within the bulk gap, which is consistent with the calculated Chern invariant.
These prove that monolayer  $\mathrm{Fe_2IX}$ (X=Cl and Br) is still QAH insulator, which is constructed from QAH insulator $\mathrm{Fe_2I_2}$\cite{fe}.

\begin{figure*}
  \includegraphics[width=12cm]{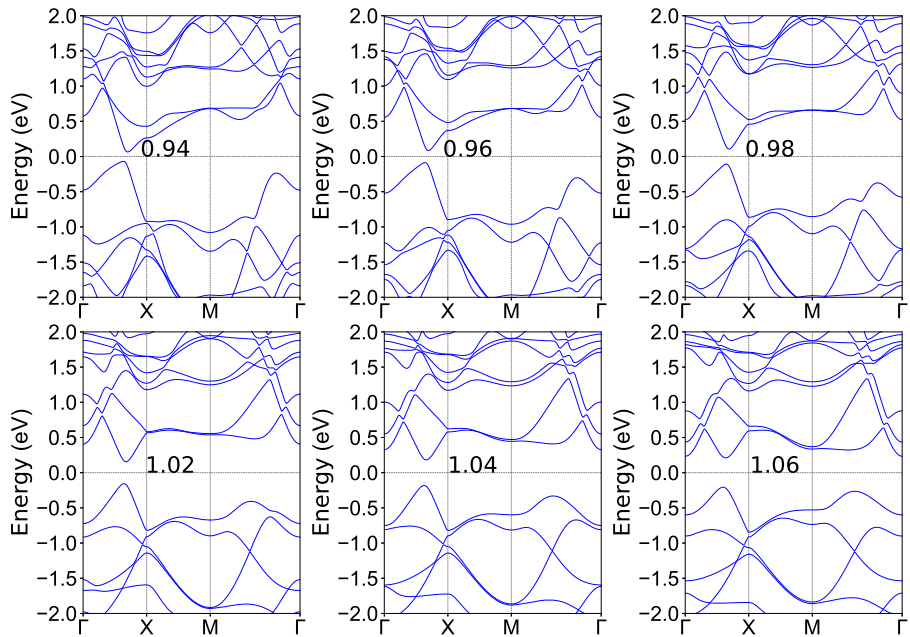}
\caption{(Color online) The energy band structures  of $\mathrm{Fe_2IBr}$ monolayer using GGA+SOC with $a/a_0$ changing from 0.94 to 1.06.}\label{t2-s}
\end{figure*}

\begin{figure}
   \includegraphics[width=7.0cm]{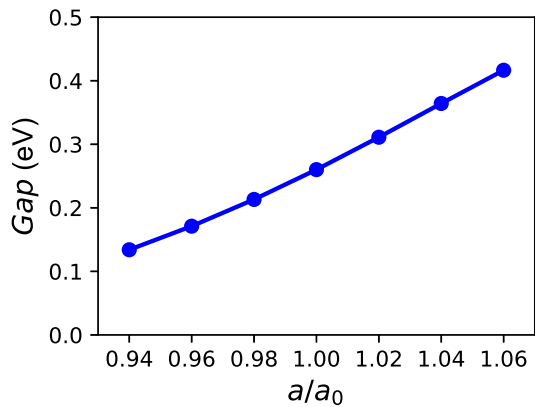}
  \caption{(Color online) For monolayer  $\mathrm{Fe_2IBr}$, the gap with the application of  biaxial strain (0.94 to 1.06) using GGA+SOC.}\label{t3-s}
\end{figure}
\begin{figure*}
   \includegraphics[width=12.0cm]{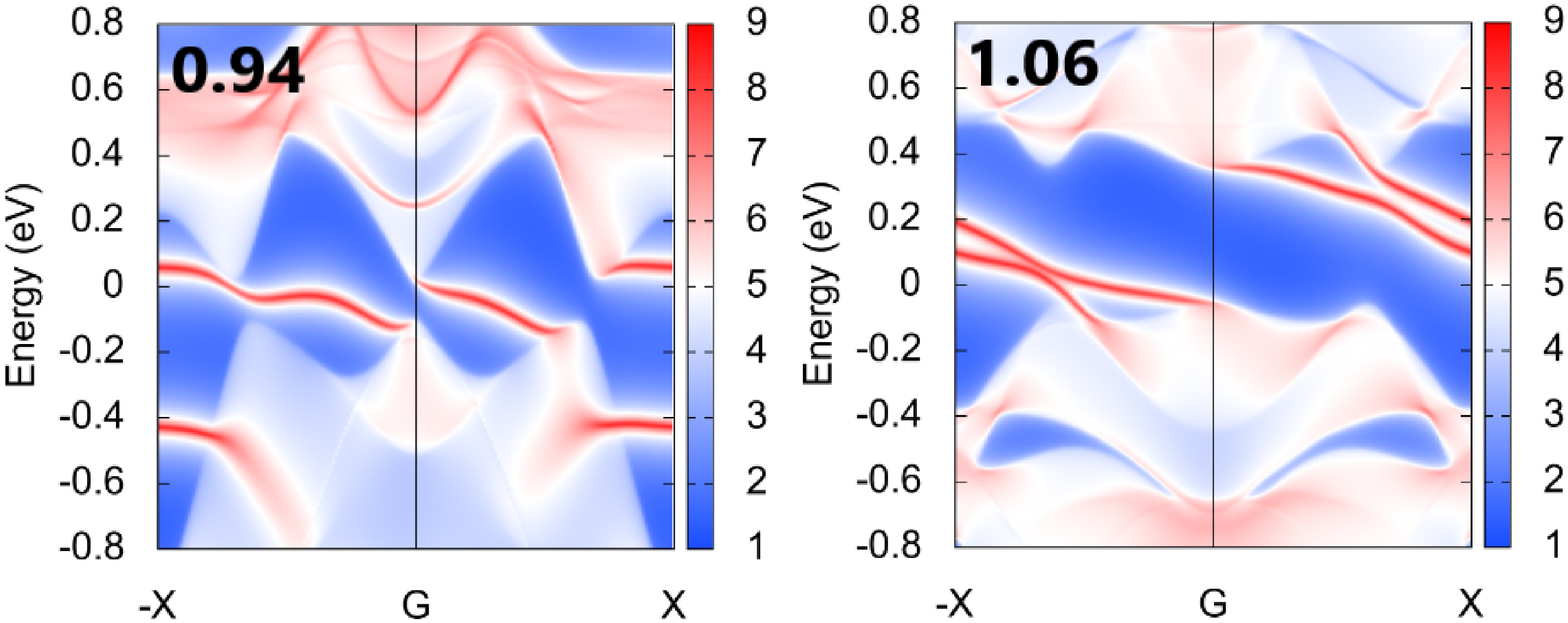}
   \includegraphics[width=12.0cm]{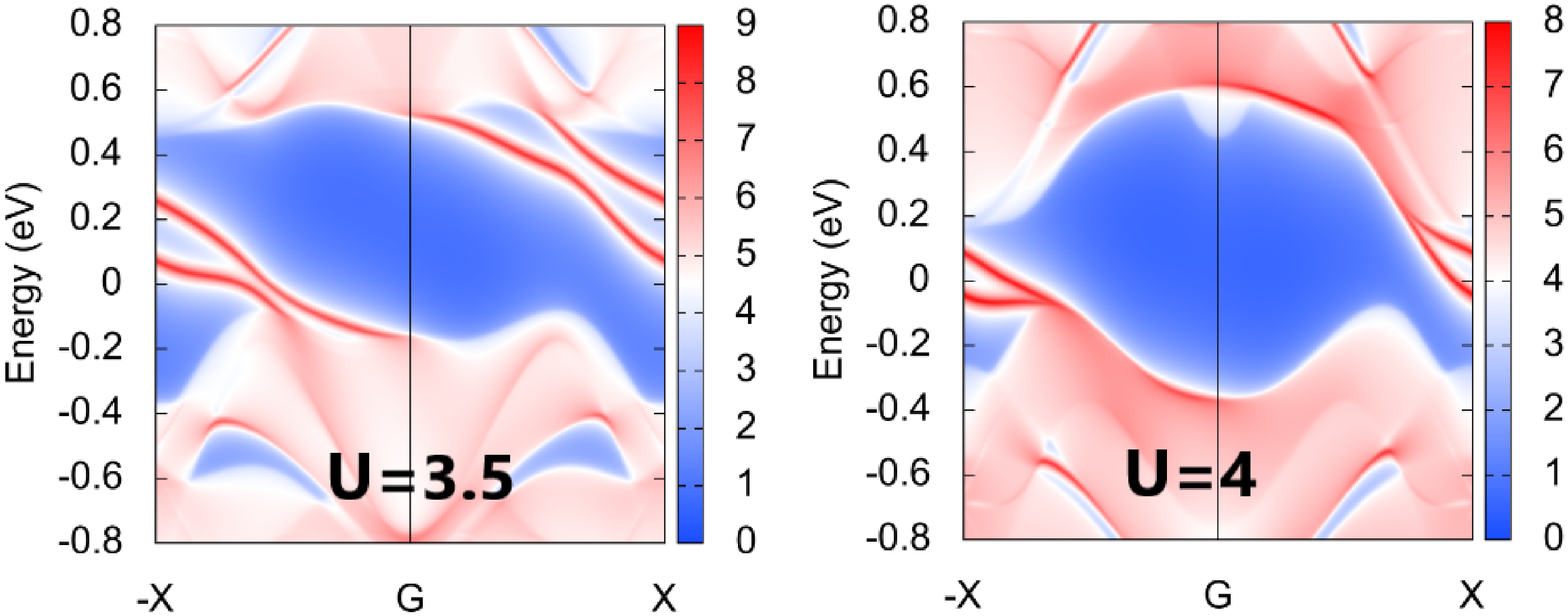}
  \caption{(Color online)Top: the topological
edge states of $\mathrm{Fe_2IBr}$  calculated  along the (100) direction at 0.94 and 1.06 strains. Bottom:the topological
edge states of $\mathrm{Fe_2IBr}$  calculated  along the (100) direction at $U$=3.5 eV and $U$=4.0 eV.}\label{t4-s}
\end{figure*}

Next, we investigate biaxial  strain effects on
the QAH properties of $\mathrm{Fe_2IX}$ (X=Cl and Br). Here, the $a/a_0$ (0.94-1.06) is used to simulate the biaxial strain, in which   $a$ and $a_0$ are  the strained and  unstrained lattice constants, respectively. From \autoref{e}, it is found that the ground states of all strained monolayers are
FM in considered strain range. The energy band structures  of $\mathrm{Fe_2IBr}$  monolayer using GGA+SOC with FM order  ($a/a_0$ from 0.94 to 1.06) are plotted in \autoref{t2-s}. It is clearly seen that all considered strained monolayers are FM semiconductors.  The gaps of monolayer $\mathrm{Fe_2IBr}$ as a function of strain are shown in \autoref{t3-s}.  The gap of $\mathrm{Fe_2IBr}$  is found to increase almost linearly with strain from 0.94 to 1.06. At 1.06 strain, the gap is up to 0.417 eV   for monolayer $\mathrm{Fe_2IBr}$. In considered strain range, the topological
edge states of monolayer $\mathrm{Fe_2IBr}$  are  calculated, and we  show  topological
edge states  at representative 0.94 and 1.06 strains in \autoref{t4-s}.
It is clearly seen that  two chiral topologically
protected gapless edge states emerge, giving  Chern number C=2.  For monolayer $\mathrm{Fe_2ICl}$, similar results can be found, and the strained energy band structures, gaps and   topological
edge states are plotted in Fig.3, Fig.4 and Fig.5 of ESI.  For monolayer $\mathrm{Fe_2ICl}$, the gap is up to 0.353 eV at 1.06 strain.
These prove  that monolayer $\mathrm{Fe_2IX}$ (X=Cl and Br) possesses robust QAH states against strain.

\section{Piezoelectric properties}
Due to   $P4/nmm$  point-group symmetry, the $\mathrm{Fe_2I_2}$ monolayer   are centrosymmetric without piezoelectricity.
 The monolayer $\mathrm{Fe_2IX}$ (X=Cl and Br) with $P4mm$ point-group symmetry lacks   reflectional
symmetry across the xy plane, but has the  reflectional
symmetry across the xz or yz plane. These mean that in-plane piezoelectricity will disappear, and only out-of-plane piezoelectricity can exit.
The third-rank piezoelectric stress tensor  $e_{ijk}$ and strain tensor $d_{ijk}$ can be used to describe
the piezoelectric effects of a material, which include  ionic
and electronic contributions.   The $e_{ijk}$ and $d_{ijk}$ are defined as:
 \begin{equation}\label{pe0}
      e_{ijk}=\frac{\partial P_i}{\partial \varepsilon_{jk}}=e_{ijk}^{elc}+e_{ijk}^{ion}
 \end{equation}
and
 \begin{equation}\label{pe0-1}
   d_{ijk}=\frac{\partial P_i}{\partial \sigma_{jk}}=d_{ijk}^{elc}+d_{ijk}^{ion}
 \end{equation}
where $P_i$, $\varepsilon_{jk}$ and $\sigma_{jk}$ are polarization vector, strain and stress, respectively.
The $e_{ijk}^{elc}$/$d_{ijk}^{elc}$ only considers electronic contributions, namely  clamped-ion piezoelectric coefficients.  The  $e_{ijk}$/$d_{ijk}$ is from the sum of ionic
and electronic contributions, neamely relax-ion piezoelectric coefficients as a realistic result.
The $e_{ijk}$ can be related with $d_{ijk}$ by elastic tensor $C_{mnjk}$:
 \begin{equation}\label{pe9-1-1}
    e_{ijk}=\frac{\partial P_i}{\partial \varepsilon_{jk}}=\frac{\partial P_i}{\partial \sigma_{mn}}.\frac{\partial \sigma_{mn}}{\partial\varepsilon_{jk}}=d_{imn}C_{mnjk}
 \end{equation}

For 2D materials, only the in-plane strain and stress ($\varepsilon_{jk}$=$\sigma_{ij}$=0 for i=3 or j=3) are taken into account\cite{q7-0,q7-1,q7-2,q7-4,q7-7,q7-8}.  By using  Voigt notation,    the  piezoelectric stress   and strain tensors of monolayer $\mathrm{Fe_2IX}$ (X=Cl and Br) can be reduced into:
 \begin{equation}\label{pe1-1}
 e=\left(
    \begin{array}{ccc}
     0 & 0 & 0 \\
     0 & 0 & 0 \\
      e_{31} & e_{31} & 0 \\
    \end{array}
  \right)
    \end{equation}

  \begin{equation}\label{pe1-2}
  d= \left(
    \begin{array}{ccc}
      0 & 0 & 0 \\
       0 & 0 & 0 \\
      d_{31} & d_{31} &0 \\
    \end{array}
  \right)
\end{equation}
With a applied uniaxial or biaxial in-plane strain,  only vertical piezoelectric polarization ($e_{31}$/$d_{31}$$\neq$0) can be produced. The $e_{31}$ can be calculated by DFPT, and the
$d_{31}$  can be  derived by \autoref{pe9-1-1}, \autoref{pe1-1} and \autoref{pe1-2}.
\begin{equation}\label{pe2}
    d_{31}=\frac{e_{31}}{C_{11}+C_{12}}
\end{equation}

\begin{table}
\centering \caption{For Janus monolayer  $\mathrm{Fe_2IX}$ (X=Cl and Br),   the elastic constants $C_{ij}$ in $\mathrm{Nm^{-1}}$, the piezoelectric stress coefficients $e_{31}$ in $10^{-10}$ C/m, the piezoelectric strain coefficients $d_{31}$ in pm/V, MAE in $\mu eV$/Fe,  easy axis (EA) and normalized exchange parameter $J$. }\label{tab0}
  \begin{tabular*}{0.48\textwidth}{@{\extracolsep{\fill}}ccccccccc}
  \hline\hline
Name& $C_{11}$& $C_{12}$&$C_{66}$ & $e_{31}$ &$d_{31}$ &MAE &EA&$J$\\\hline
$\mathrm{Fe_2ICl}$& 49.28&16.07&23.68&0.305&0.467&39&$c$&57.0 \\\hline
$\mathrm{Fe_2IBr}$&48.53&13.27&20.91&0.237&0.384&318&$c$&53.5 \\\hline\hline
\end{tabular*}
\end{table}

\begin{figure*}
   \includegraphics[width=12cm]{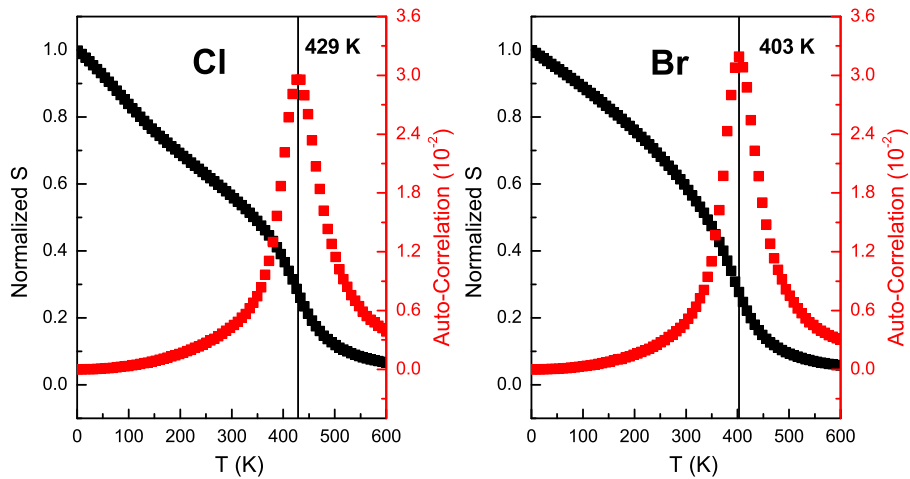}
    \caption{(Color online)The normalized magnetic moment (S) and auto-correlation of Janus monolayer  $\mathrm{Fe_2IX}$ (X=Cl and Br) as a function of temperature.  }\label{ed}
\end{figure*}
\begin{figure}
   \includegraphics[width=8.0cm]{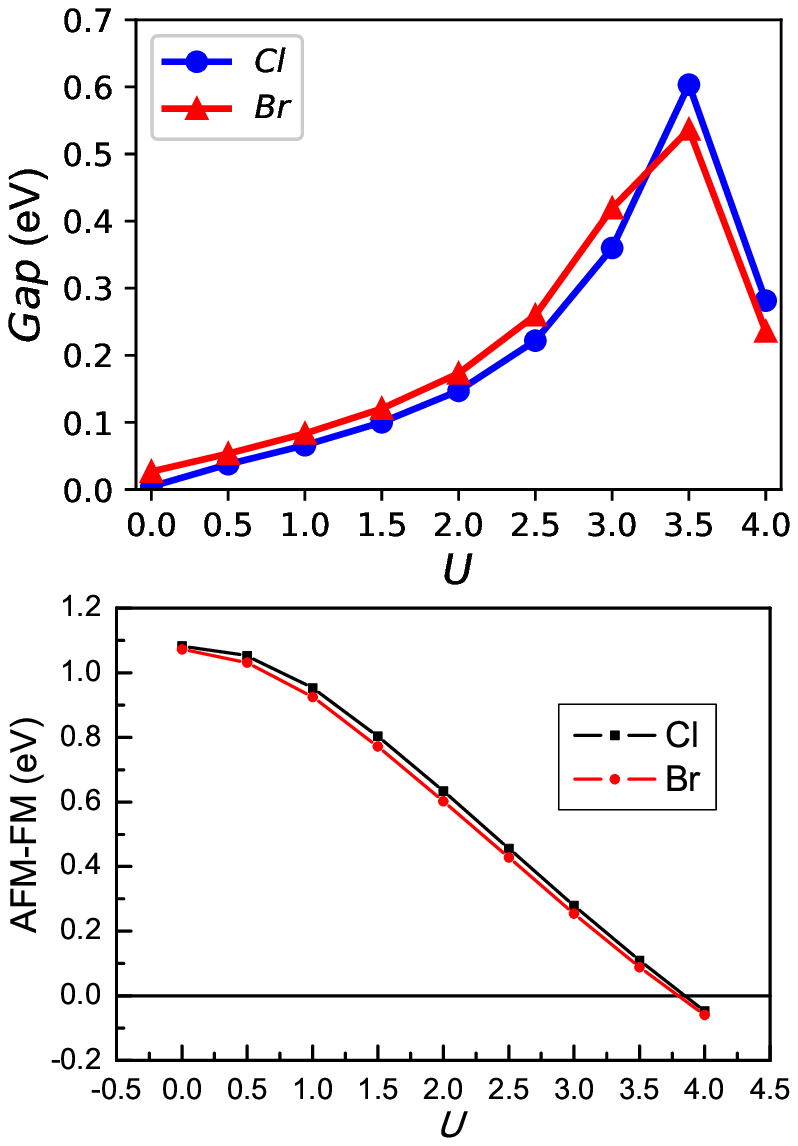}
  \caption{(Color online) For monolayer  $\mathrm{Fe_2IX}$ (X=Cl and Br), Top: the gap as a function of $U$ (0.0 to 4.0 eV) using GGA+SOC. Bottom:the energy difference between AFM and FM states as a function of $U$ using GGA.}\label{t3-s-1}
\end{figure}

The primitive cell is used  to calculate the  $e_{31}$  of  Janus monolayer  $\mathrm{Fe_2IX}$ (X=Cl and Br).
The calculated $e_{31}$ is 0.305$\times$$10^{-10}$ C/m  with ionic part   0.027$\times$$10^{-10}$ C/m  and electronic part 0.278 $\times$$10^{-10}$ C/m for  $\mathrm{Fe_2ICl}$, and     0.237$\times$$10^{-10}$ C/m with ionic contribution   0.087$\times$$10^{-10}$ C/m  and electronic contribution 0.150$\times$$10^{-10}$ C/m for $\mathrm{Fe_2IBr}$.
Calculated results show that the electronic and
ionic polarizations  have  the same signs, and  the electronic  contribution
is larger than the ionic part for Janus monolayer  $\mathrm{Fe_2IX}$ (X=Cl and Br).
According to \autoref{pe2}, the $d_{31}$ can be attained, and the calculated value is 0.467 pm/V for $\mathrm{Fe_2ICl}$,   and 0.384 pm/V for $\mathrm{Fe_2IBr}$. To be compatible with the
nowadays bottom/top gate technologies, a large out-of-plane piezoelectric $d_{31}$ is
highly desired for 2D materials.
The $d_{31}$ of Janus monolayer  $\mathrm{Fe_2IX}$ (X=Cl and Br) is  higher  than ones of many 2D  materials, like Janus TMD monolayers (0.03 pm/V)\cite{q7-0},
functionalized h-BN (0.13 pm/V)\cite{o1} and kalium decorated graphene (0.3
pm/V)\cite{o2}, and is comparable with ones of the oxygen functionalized MXenes (0.40-0.78 pm/V)\cite{q9},  Janus group-III materials (0.46 pm/V)\cite{q7-6}, Janus BiTeI/SbTeI  monolayer (0.37-0.66 pm/V)\cite{o3} and $\alpha$-$\mathrm{In_2Se_3}$
(0.415 pm/V)\cite{o4}.

\section{Curie temperature}
Curie temperatures ($T_C$) is a very important parameter for the practical application of monolayer $\mathrm{Fe_2IX}$ (X=Cl and Br).  By using Heisenberg model, the MC simulations
 are carried out to estimate the $T_C$ of monolayer $\mathrm{Fe_2IX}$ (X=Cl and Br). The spin Heisenberg Hamiltonian is defined as:
  \begin{equation}\label{pe0-1-1}
H=-J\sum_{i,j}S_i\cdot S_j-A\sum_i(S_i^z)^2
 \end{equation}
in which  $J$, $S$ and $A$ are the nearest neighbor exchange parameter,  the
spin vector of each atom and   anisotropy energy
parameter, respectively. Using the energy
difference between AFM and FM  and normalized
$S$ ($|S|$ = 1), the magnetic coupling parameters
are calculated as $J$=($E_{AFM}$-$E_{FM}$)/8. The  calculated $J$ value is 57.0 meV for  $\mathrm{Fe_2ICl}$,  and 53.5 meV for  $\mathrm{Fe_2IBr}$.

The  magnetic anisotropy energy (MAE) plays
a very important role to determine the thermal stability of magnetic
ordering, which  mainly originates from the
SOC interaction. The GGA+SOC calculations are used to obtain relative stabilities along
the (100) and  (001) directions.  The related data are summarized in \autoref{tab0}. It is found that  the easy axes of these monolayers are along the
out-of-plane (001) direction. A  40$\times$40 supercell and  $10^7$ loops are adopted to carry out the
MC simulation. The
 normalized magnetic moment and auto-correlation of monolayer $\mathrm{Fe_2IX}$ (X=Cl and Br) as a function of temperature are plotted in \autoref{ed}.
It can be seen that $T_C$  is as
high as 429 K/403 K for  $\mathrm{Fe_2ICl}$/$\mathrm{Fe_2IBr}$,  which is significantly higher than that of previously
reported many  2D FM semiconductors, like $\mathrm{CrI_3}$ monolayer (about 45 K)\cite{m7-6}, CrOCl monolayer
(about 160 K)\cite{tc1} and $\mathrm{Cr_2Ge_2Te_6}$ monolayer (about 20 K)\cite{tc2}.  The  $T_C$ for monolayer  $\mathrm{Fe_2ICl}$/$\mathrm{Fe_2IBr}$  is very close to one of $\mathrm{Fe_2I_2}$ monolayer\cite{fe}. Considering the large nontrivial
band gap and out-of-plane piezoelectric polarizations combined with the high $T_C$, monolayer $\mathrm{Fe_2IX}$ (X=Cl and Br)
would provide a promising platform for exploring the
PQAH effect at room temperature.

\section{Discussion and Conclusion}
To understand the relationship between electronic correlation
and nontrivial band topology, the electronic structures of monolayer $\mathrm{Fe_2IX}$ (X=Cl and Br)  are calculated by GGA+SOC with $U$ from 0.0 eV to 4.0 eV. Some representative energy bands of monolayer $\mathrm{Fe_2IX}$ (X=Cl and Br) are plotted in Fig.6 and Fig.7 of ESI, and
their gaps as a function of $U$ (0.0 to 4.0 eV) are shown in \autoref{t3-s-1}. With increasing $U$, it is clearly seen that the gap  increases firstly, and then decreases. At $U$=0.0 eV, the gap is 4 meV for $\mathrm{Fe_2ICl}$, and 27 meV for $\mathrm{Fe_2IBr}$.  At about $U$=3.5 eV, their gaps reach  maximums, and 604 meV for $\mathrm{Fe_2ICl}$  and 537 meV  for $\mathrm{Fe_2IBr}$. At $U$=4 eV, their gaps become small, which is because the  conduction band minimum (CBM)/valence band maximum (VBM)  changes from one point along $\Gamma$-X line to M point/one point along $\Gamma$-M line.
In considered $U$ range, the topological
edge states of monolayer $\mathrm{Fe_2IX}$ (X=Cl and Br)  are  calculated. We  show  topological
edge states of $\mathrm{Fe_2IBr}$  at representative $U$=3.5  and 4 eV in \autoref{t4-s}, and in Fig.8 of ESI for $\mathrm{Fe_2ICl}$.
It is clearly seen that  two chiral topologically
protected gapless edge states emerge, which means that the  Chern number C=2.
These show  that monolayer $\mathrm{Fe_2IX}$ (X=Cl and Br) possesses robust QAH states against electronic correlation.
Finally, the energy difference between AFM and FM states as a function of $U$ for monolayer $\mathrm{Fe_2IX}$ (X=Cl and Br) are plotted in \autoref{t3-s-1}.
It is found that the ground state of monolayer $\mathrm{Fe_2IX}$ (X=Cl and Br) becomes AFM, when the $U$ is larger than about 3.75 eV.

In summary,  the  intriguing 2D  PQAHIs $\mathrm{Fe_2IX}$ (X=Cl and Br) are predicted by the reliable first-principle calculations.
Monolayer  $\mathrm{Fe_2IX}$ (X=Cl and Br) exhibits excellent
dynamic, mechanical and thermal stabilities, and possesses out-of-plane magnetic anisotropy and
high Curie temperature. The nontrivial
topology of $\mathrm{Fe_2IX}$ (X=Cl and Br), identified with  Chern
number C=2 and chiral edge states, has a nontrivial
band gap of more than 200 meV.  The calculated  out-of-plane $d_{31}$   is  higher than  or comparable with ones of  other 2D known materials, which is highly desirable for ultrathin piezoelectric devices. It is proved that  the emergence of
robust QAH states in monolayer $\mathrm{Fe_2IX}$ (X=Cl and Br) is robust against strain and electronic correlation.
Our predicted  room-temperature PQAHIs are of crucial
importance to fundamental research and to future development
of electronics, piezoelectronics and  spintronics, and these findings open
new opportunities to realize novel practical quantum
applications.

~~~~\\
~~~~\\
\textbf{Conflicts of interest}
\\
There are no conflicts to declare.

\begin{acknowledgments}
This work is supported by Natural Science Basis Research Plan in Shaanxi Province of China  (2021JM-456). We are grateful to the Advanced Analysis and Computation Center of China University of Mining and Technology (CUMT) for the award of CPU hours and WIEN2k/VASP software to accomplish this work.
\end{acknowledgments}


\begin{references}

\bibitem{z}P. Lin, C. Pan and  Z. L. Wang, Materials Today Nano \textbf{4}, 17 (2018).

\bibitem{q5}M. Dai, Z. Wang, F. Wang, Y. Qiu, J. Zhang, C. Y. Xu, T. Zhai, W. Cao, Y. Fu,
D. Jia, Y. Zhou, and P. A. Hu, Nano Lett. \textbf{19}, 5416 (2019).

\bibitem{q6} W. Wu, L. Wang, Y. Li, F. Zhang, L. Lin, S. Niu, D. Chenet,
X. Zhang, Y. Hao, T. F. Heinz, J. Hone and Z. L. Wang,
Nature \textbf{514}, 470 (2014).

\bibitem{q8}A. Y. Lu, H. Zhu, J. Xiao, C. P. Chuu, Y. Han, M. H. Chiu,
C. C. Cheng, C. W. Yang, K. H. Wei, Y. Yang, Y. Wang,
D. Sokaras, D. Nordlund, P. Yang, D. A. Muller, M. Y. Chou,
X. Zhang and L. J. Li, Nat. Nanotechnol. \textbf{12}, 744 (2017).

\bibitem{q8-1}H. Zhu, Y. Wang, J. Xiao, M. Liu, S. Xiong, Z. J. Wong, Z. Ye,
Y. Ye, X. Yin and X. Zhang, Nat. Nanotechnol. \textbf{10},
151 (2015).


\bibitem{q7-0}L. Dong, J. Lou and V. B. Shenoy, ACS Nano, \textbf{11},
8242 (2017).

\bibitem{q7-1}R. X. Fei, We. B. Li, J. Li and L. Yang, Appl. Phys. Lett.  \textbf{107}, 173104 (2015).



\bibitem{q7-2}M. N. Blonsky, H. L. Zhuang, A. K. Singh and R.  G. Hennig,  ACS Nano, \textbf{9},
9885 (2015).

\bibitem{q7-4} S. D. Guo, Y. T. Zhu, W. Q. Mu and W. C. Ren,  EPL \textbf{132},  57002 (2020).


\bibitem{q7-7}W. B. Li  and J. Li, Nano Res.  \textbf{8}, 3796 (2015).


\bibitem{q7-8}Dimple, N. Jena, A. Rawat, R.  Ahammed,
M. K. Mohanta and A. D. Sarkar, J. Mater. Chem. A  \textbf{6},
24885 (2018).



\bibitem{q7-10}N. Jena, Dimple, S. D.  Behere  and A. D. Sarkar, J. Phys. Chem. C  \textbf{121}, 9181 (2017).

\bibitem{q9-0}M. T. Ong and E.J. Reed,  ACS Nano \textbf{6}, 1387 (2012).

\bibitem{q9-1}A. A. M. Noor, H. J. Kim  and Y. H. Shin, Phys. Chem. Chem. Phys. \textbf{16}, 6575 (2014).


\bibitem{q9}J. Tan, Y. H. Wang, Z. T. Wang, X. J. He, Y. L. Liu, B. Wanga, M. I. Katsnelson and  S. J.  Yuan, Nano Energy \textbf{65},  104058 (2019).

\bibitem{qt1}J. H. Yang,  A. P. Wang, S. Z. Zhang, J.  Liu, Z. C. Zhong and L. Chen, Phys. Chem. Chem. Phys.,
\textbf{21}, 132 (2019).

\bibitem{q15}S. D. Guo, W. Q. Mu, Y. T. Zhu and X. Q. Chen, Phys. Chem. Chem. Phys. \textbf{22}, 28359 (2020).

\bibitem{q15-1}S. D. Guo, X. S. Guo, X. X. Cai, W. Q. Mu and  W. C. Ren, 	arXiv:2103.15141 (2021).

\bibitem{q15-2}G. Song, D. S. Li, H. F. Zhou et al., Appl. Phys. Lett. \textbf{118}, 123102 (2021).


\bibitem{gsd1}S. D. Guo, W. Q. Mu, Y. T. Zhu, S. Q. Wang and  G. Z. Wang, 	J. Mater. Chem. C \textbf{9}, 5460 (2021).

\bibitem{gsd2}S. D. Guo, Y. T. Zhu, W. Q. Mu and X. Q. Chen, arXiv:2103.03456 (2021).


\bibitem{m7-1} Y. Ma, Y. Dai, M. Guo, C. Niu, Y. Zhu and B. Huang, ACS
Nano, \textbf{6}, 1695 (2012).

\bibitem{m7-2} C. Gong, L. Li, Z. Li, H. Ji, A. Stern, Y. Xia, T. Cao, W. Bao,
C. Wang, Y. Wang, Z. Q. Qiu, R. J. Cava, S. G. Louie, J. Xia
and X. Zhang, Nature \textbf{546}, 265 (2017).

\bibitem{m7-3}M. Khazaei,  M. Arai,  T. Sasaki,  C. Y. Chung,  N. S. Venkataramanan,  M. Estili,  Y. Sakka and   Y. Kawazoe, Adv. Funct. Mater. \textbf{23}, 2185 (2013).


\bibitem{m7-4} Y. Guo, H. Deng, X. Sun, X. Li, J. Zhao, J. Wu, W. Chu,
S. Zhang, H. Pan, X. Zheng, X. Wu, C. Jin, C. Wu and Y. Xie,
Adv. Mater. \textbf{29}, 1700715 (2017).

\bibitem{m7-6}B. Huang, G. Clark, E. Navarro-Moratalla, D. R. Klein, R. Cheng,
K. L. Seyler, D. Zhong, E. Schmidgall, M. A. McGuire, D. H.
Cobden, W. Yao, D. Xiao, P. Jarillo-Herrero and X. Xu, Nature \textbf{546}, 270 (2017).



\bibitem{m7-7}G. Bhattacharyya,    I. Choudhuri,   P. Bhauriyal,  P. Garg   and  B. Pathak, Nanoscale \textbf{10}, 22280 (2018).


\bibitem{m7-8}W. B. Zhang,  Q.  Qu,  P. Zhu and C. H. Lam, : J. Mater. Chem. C  \textbf{3}, 12457 (2015).


\bibitem{qa1}C. Z. Chang, J. S. Zhang,  X. Feng  et al.,  Science   \textbf{340}, 167 (2013).

\bibitem{qa2}J. G. Checkelsky,  J. T. Ye,  Y. Onose et al., Nat. Phys.   \textbf{8}, 729 (2012).

\bibitem{fe}Q. L. Sun,  Y. D. Ma and N. Kioussis, Mater. Horiz. \textbf{7}, 2071 (2020).


\bibitem{p1}A. Y. Lu, H. Y. Zhu, J. Xiao et al., Nature Nanotechnology \textbf{12}, 744 (2017).

\bibitem{1}P. Hohenberg and W. Kohn, Phys. Rev. \textbf{136},
B864 (1964); W. Kohn and L. J. Sham, Phys. Rev. \textbf{140},
A1133 (1965).

\bibitem{pv1} G. Kresse, J. Non-Cryst. Solids \textbf{193}, 222 (1995).

\bibitem{pv2} G. Kresse and J. Furthm$\ddot{u}$ller, Comput. Mater. Sci. 6, \textbf{15} (1996).

\bibitem{pv3} G. Kresse and D. Joubert, Phys. Rev. B \textbf{59}, 1758 (1999).
\bibitem{pbe}J. P. Perdew, K. Burke and M. Ernzerhof, Phys. Rev. Lett. \textbf{77}, 3865 (1996).

\bibitem{u}V. I. Anisimov, F. Aryasetiawan and A. I. Lichtenstein,
J. Phys. Condens. Mat. \textbf{9}, 767 (1997).

\bibitem{fe1}J. Heyd, G. E. Scuseria and M. Ernzerhof, J. Chem. Phys. \textbf{118}, 8207 (2003).


\bibitem{pv6}X. Wu, D. Vanderbilt and  D. R.  Hamann, Phys. Rev. B  \textbf{72}, 035105 (2005).


\bibitem{pv5}A. Togo, F. Oba, and I. Tanaka, Phys. Rev. B \textbf{78}, 134106
(2008).

\bibitem{w1} Q. Wu, S. Zhang, H. F. Song, M. Troyer and A. A. Soluyanov, Comput. Phys. Commun. \textbf{224}, 405
(2018).
\bibitem{w2}A. A. Mostofia, J. R. Yatesb, G. Pizzif, Y.-S. Lee, I. Souzad, D.
Vanderbilte and N. Marzarif,  Comput. Phys. Commun. \textbf{185}, 2309 (2014).


\bibitem{mc}L. Liu, X. Ren, J. H. Xie, B. Cheng, W. K. Liu, T. Y. An, H. W. Qin and  J. F. Hu, Appl. Surf. Sci.  \textbf{480},  300 (2019).




\bibitem{ela}E. Cadelano and L. Colombo, Phys. Rev. B  \textbf{85}, 245434 (2012).

\bibitem{ela1} E. Cadelano, P. L. Palla, S. Giordano, and L. Colombo, Phys.
Rev. B  \textbf{82}, 235414 (2010).

\bibitem{gra}C. Lee, X. Wei, J. W. Kysar, and J. Hone, Science \textbf{321}, 385
(2008).



\bibitem{pr}M. Eremin and Y. V. Rakitin, J. Phys. C \textbf{14}, 247 (1981).

\bibitem{pr1}Y. Li, J. H. Li, Y. Li et al., Phys. Rev. Lett. \textbf{125}, 086401 (2020).

\bibitem{o1}A. A. M. Noor, H. J. Kim and Y. H. Shin,  Phys. Chem.
Chem. Phys. \textbf{16}, 6575 (2014).

\bibitem{o2} M. T. Ong and E. J. Reed, ACS Nano \textbf{6},  1387 (2012).

\bibitem{q7-6}Y. Guo, S. Zhou, Y. Z. Bai, and J. J. Zhao, Appl. Phys. Lett. \textbf{110}, 163102 (2017).

\bibitem{o3}S. D. Guo, X. S. Guo, Z. Y. Liu and Y. N. Quan,  J. Appl. Phys. \textbf{127}, 064302 (2020).


\bibitem{o4}L. Hu and X.R. Huang, RSC Adv. \textbf{7},  55034 (2017).



\bibitem{tc1}N. Miao, B. Xu, L. Zhu, J. Zhou and Z. Sun, J. Am. Chem. Soc. \textbf{140}, 2417 (2018).


\bibitem{tc2}C. Gong, L. Li, Z. Li, H. Ji, A. Stern, Y. Xia, T. Cao, W. Bao,
C. Wang, Y. Wang, Z. Q. Qiu, R. J. Cava, S. G. Louie, J. Xia
and X. Zhang, Nature  \textbf{546}, 265 (2017).

\end{references}
\end{document}